\documentclass[%
reprint,
superscriptaddress,
nofootinbib,
 amsmath,amssymb,
 aps,
]{revtex4-1}

\usepackage{graphicx}
\usepackage{dcolumn}
\usepackage{bm}

\newcommand\apjl{ApJL}%
          
\begin{document}

\preprint{APS/123-QED}

\title{Extracting the orbital axis from gravitational waves of precessing binary systems}

\author{Kyohei Kawaguchi}
\affiliation{Max Planck Institute for Gravitational Physics (Albert Einstein Institute), Am M\"{u}hlenberg 1, Potsdam-Golm, 14476, Germany}
\affiliation{Center for Gravitational Physics, Yukawa Institute for Theoretical Physics, Kyoto University, Kyoto 606-8502, Japan}
\author{Koutarou Kyutoku}
\affiliation{Theory Center, Institute of Particle and Nuclear Studies, KEK, Tsukuba 305-0801, Japan}
\affiliation{Department of Particle and Nuclear Physics, the Graduate University for Advanced Studies (Sokendai), Tsukuba 305-0801, Japan}
\affiliation{Interdisciplinary Theoretical Science (iTHES) Research Group, RIKEN, Wako, Saitama 351-0198, Japan}
\affiliation{Center for Gravitational Physics, Yukawa Institute for Theoretical Physics, Kyoto University, Kyoto 606-8502, Japan}
\author{Hiroyuki Nakano}
\affiliation{Faculty of Law, Ryukoku University, 67 Fukakusa Tsukamoto-cho, Fushimi-ku, Kyoto 612-8577, Japan}
\affiliation{Center for Gravitational Physics, Yukawa Institute for Theoretical Physics, Kyoto University, Kyoto 606-8502, Japan}
\author{Masaru Shibata}
\affiliation{Center for Gravitational Physics, Yukawa Institute for Theoretical Physics, Kyoto University, Kyoto 606-8502, Japan}
\date{\today}

\begin{abstract}
	We present a new method for extracting the instantaneous orbital axis only from gravitational wave strains of precessing binary systems observed from a particular observer direction. This method enables us to reconstruct the co-precessing frame waveforms only from observed strains for the ideal case with the high signal-to-noise ratio. Specifically, we do not presuppose any theoretical model of the precession dynamics and co-precessing waveforms in our method. We test and measure the accuracy of our method using the numerical relativity simulation data of precessing binary black holes taken from the SXS Catalog. We show that the direction of the orbital axis is extracted within $\approx0.07~{\rm rad}$ error from gravitational waves emitted during the inspiral phase. The co-precessing waveforms are also reconstructed with high accuracy; the mismatch (assuming white noise) between them and the original co-precessing waveforms is typically a few times $10^{-3}$ including the merger-ringdown phase, and can be improved by an order of magnitude focusing only on the inspiral waveform. In this method, the co-precessing frame waveforms are not only the purely technical tools for understanding the complex nature of precessing waveforms but also direct observables.
\begin{description}
\item[PACS numbers]04.30.−w, 04.25.dg, 
\end{description}
\end{abstract}
\maketitle


\section{Introduction}
	The three detections (and one candidate) of gravitational waves from binary black hole mergers achieved by Advanced LIGO~\citep{2016PhRvL.116m1103A,2016PhRvL.116x1103A,2016arXiv160604856T,2017PhRvL.118v1101A} have marked the beginning of the era of gravitational-wave astronomy. In particular, the first detection was achieved with significantly large signal-to-noise ratio.  The information provided by the gravitational-wave observation surely enhances our knowledge about the universe, and in particular, the black holes. A number of merger events will be detected by Advanced LIGO and the following running of the other ground-based detectors, such as VIRGO~\citep{2015CQGra..32b4001A}, and KAGRA~\citep{2012CQGra..29l4007S}.  In addition, third-generation ground-base detectors, for which the sensitivity is by an order of magnitude higher than the current detectors, are proposed~\citep{Punturo2010}. Furthermore, space-based detectors, such as LISA~\citep{2017arXiv170200786A} and DECIGO~\citep{2011CQGra..28i4011K}, will be powerful observatories to detect massive binary black holes with significantly high signal-to-noise ratios.
	
	If either of the directions of two black-hole spins is not aligned with the orbital axis, the orbital precession occurs in the inspiral phase of the binary coalescence~\citep{1994PhRvD..49.6274A,1995PhRvD..52..821K}. Such orbital precession strongly affects the gravitational waveforms by modulating both amplitude and phase. The complex nature of the waveforms from a precessing binary contains richer information about the binary parameters than without the orbital precession~\citep{2011PhRvD..84b2002L,2013PhRvD..88f2001A,2013PhRvD..87b4004C,2009PhRvD..80f4027K}. However, the complexity also makes it difficult to understand the dependence of waveforms on the parameters.
		
	Many efforts have been made to model precessing waveforms, and many frameworks have been developed to simplify those complex features~\citep{2003PhRvD..67j4025B,2014PhRvD..89f1502T,2014PhRvD..89h4006P,2017PhRvD..95b4010B,2014PhRvL.113o1101H,2017arXiv170100550B,2011PhRvD..84l4011B,2011PhRvD..84l4002O,2011PhRvD..84b4046S,2012PhRvD..86j4063S,2014arXiv1409.4431B}. Most remarkably, in Refs.~\citep{2003PhRvD..67j4025B,2011PhRvD..84l4002O,2011PhRvD..84b4046S,2012PhRvD..86j4063S,2011PhRvD..84l4011B}, it is shown that the inspiral waveforms from a precessing binary can be dramatically simplified in the so-called ``co-precessing frame'', which follows the instantaneous orbital plane of the binary. The precessing waveforms in such a frame become just as if they are from a non-precessing binary. Also, the approximate mapping between the precessing waveforms and non-precessing waveforms has been proposed~\citep{2012PhRvD..86j4063S}. Working in the co-precessing frame enables us to understand and to model the waveforms from precessing binaries much more easily than working in the inertial frame. We note that, even for the case that the binary is not precessing, the modulation arises for an observer due to the mode coupling if the line of sight is misaligned with the orbital axis.
	
	However, to extract the instantaneous orbital axis and to obtain the co-precessing frame waveforms, we need knowledge of gravitational waveforms observed from all the directions, or at least, $l=2$ components of the spherical harmonics in the inertial frame. On the other hand, we can only obtain the strain from a particular observer direction in real observations. Therefore, it is difficult to apply the framework of co-precessing frame directly for the observation, and hence, the co-precessing waveforms have only been treated as the intermediates for modeling the waveforms in the inertial frame.

	In this paper, we present a new method for extracting the instantaneous orbital axis and for reconstructing the co-precessing frame waveforms only from gravitational wave strains observed from a particular observer direction. To introduce our method and to show that the systematic error associated with our method is acceptably small, as a first step, we assume the case that the detector noise is negligible to analyze the waveforms directly. We test and measure the accuracy of our method using the numerical relativity simulation data of the precessing binary black holes taken from SXS Catalog~\citep{SXS:catalog,Mroue:2012kv,Mroue:2013xna,Hinder:2013oqa}. Our analysis does not presuppose any theoretical model of the precession dynamics and co-precessing waveforms, and thus, the method can also be used for the case that the time evolution does not obey the prediction of general relativity. Our method is composed of two basic ideas: One is the transformation, which we call the mode decomposition, that decomposes the wave strain into Fourier(-like) components in terms of the harmonic modes in the co-precessing frame rather than the frequency. The other is the procedure to extract the ``orbital phase'' of the binary for use in the mode decomposition only from the precessing wave strain.
	
	Before moving to the explanation of our method, we summarize conventions and basic assumption which we employ in this paper. Throughout this paper, we employ the geometrical units $c=G=1$, where $c$ and $G$ are the speed of light and the gravitational constant, respectively. We refer to the total mass of the system at the infinite separation as $M$. Among several definitions for the co-precessing frame~\citep{2003PhRvD..67j4025B,2011PhRvD..84l4002O,2011PhRvD..84b4046S,2012PhRvD..86j4063S}, in this paper, we employ the so-called quadrupole-preferred frame (referred to as the quadrupole-aligned (QA) frame or the QA method in the following) introduced in Refs.~\citep{2011PhRvD..84b4046S,2012PhRvD..86j4063S}. We refer to $z$-axis obtained by the quadrupole-preferred frame as the direction of the orbital angular momentum or the instantaneous orbital axis, ${\hat {\bf L}}$ ($|{\hat {\bf L}}|=1$), just for simplicity. We note that ${\hat {\bf L}}$ does not always agree with and rather slightly deviates from the Newtonian orbital angular momentum, ${\hat {\bf L}}_{\rm N}$, defined in Refs.~\citep{1995PhRvD..52..821K,2003PhRvD..67j4025B} due to the higher order post-Newtonian corrections.

\begin{figure}
	\begin{center}
		\includegraphics[width=80mm]{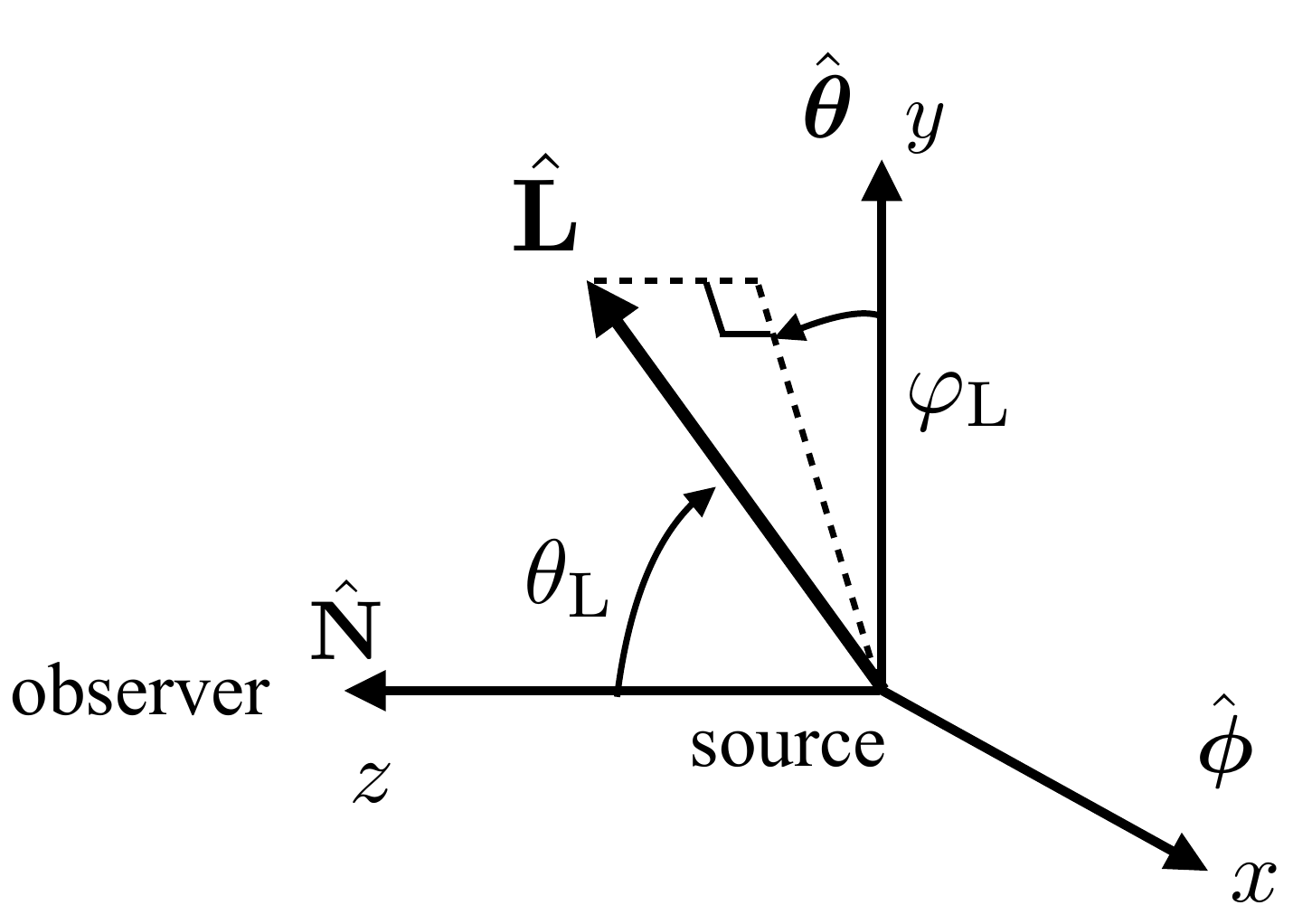}
	\end{center}
	\caption{The definitions of the angles in the source frame. The unit vector, ${\hat {\bf N}}$, denotes the direction from the source to the observer. ${\hat {\boldsymbol \theta}}$ and ${\hat {\boldsymbol \phi}}$ denote unit vectors in the directions of $(\partial/\partial\theta)^i$ and $(\partial/\partial\phi)^i$, respectively, where those two angles describe the sky position of the source.}
	\label{fig:def_ang}
\end{figure}
	To describe the precession of the binary, we introduce a coordinate system as follows. First, we describe the direction of the source in the sky by two polar angles, $\theta$ and $\phi$, and define a unit vector, ${\hat {\bf N}}$, as the direction from the source to the observer. Next, we introduce two bases in the plane perpendicular to ${\hat {\bf N}}$, ${\hat {\boldsymbol \theta}}$ and ${\hat {\boldsymbol \phi}}$, which are the unit vectors in the directions of $(\partial/\partial\theta)^i$ and $(\partial/\partial\phi)^i$, respectively. Then, we introduce a Cartesian coordinate system, $(x,y,z)$, in the source frame in such a way that $x, y$, and $z$ directions agree with ${\hat {\boldsymbol \phi}}$, ${\hat {\boldsymbol \theta}}$, and ${\hat {\bf N}}$, respectively (see Fig.~\ref{fig:def_ang}). We describe the direction of the orbital angular momentum, ${\hat {\bf L}}\left(t\right)$, by introducing two polar angles, $\theta_{\rm L}\left(t\right)$ and $\varphi_{\rm L}\left(t\right)$, in the source frame defined by
\begin{align}
	\theta_{\rm L}\left(t\right)={\rm cos}^{-1}\left[{\hat L}_z\left(t\right)\right],\\
	\varphi_{\rm L}\left(t\right)={\rm Arg}\left[{\hat L}_x\left(t\right)+i{\hat L}_x\left(t\right)\right]-\frac{\pi}{2},
\end{align}
where ${\hat L}_i~(i=x,y,z)$ is a component of ${\hat {\bf L}}$ in the source frame. We note that we shift $\varphi_{\rm L}$ by $-\pi/2$ so that ${\hat {\bf L}}$ lies in $yz$-plane for the case $\varphi_{\rm L}=0$.

	We denote a complex waveform strain by $h=h_+-ih_\times$, where $h_+$ and $h_\times$ are the plus and cross modes of gravitational-wave polarization defined by
\begin{align}
	h_+=&\frac{1}{2}\left(h^{\rm TT}_{{\hat \theta}{\hat \theta}}-h^{\rm TT}_{{\hat \phi}{\hat \phi}}\right),\\
	h_\times=&-h^{\rm TT}_{{\hat \theta}{\hat \phi}}.
\end{align}
Here,  $h^{\rm TT}_{ij}$ is a transverse-traceless component of the metric perturbation. We note that the sign of $h_\times$ is opposite from the usual definition due to our different choice of the coordinate system.

In this paper, we focus only on the case that the complex waveform strain, $h$, is known and do not consider the effect of the noise to demonstrate the usefulness of our method. Using ground-based detectors, multiple detectors are needed to determine $h$. The sky localization of the event is also important to determine $h$ accurately. The follow-up observations of electromagnetic counterparts will help for determining the sky location of the events including neutron stars~\citep{2016ARNPS..66...23F}. In the observations of binary black holes by space-based detectors such as  LISA and DECIGO, our method will be useful because they could determine the sky position accurately~\citep{2003ApJ...596L.231T,2007PhRvD..76b2003K}. We leave the study on how the errors in the observation influence the accuracy of extracting the orbital axis in our method for future study. We also note that, in this paper, our method is only tested for the data of precessing binary black holes for which the precessing time scale is always much longer than their orbital period except just before the mergers.
 
\section{Method}
\subsection{Mode decomposition}
	The waveforms from a precessing binary observed in the inertial frame can be described by using co-precessing frame waveforms as~\citep{2011PhRvD..84b4046S,2012PhRvD..86j4063S},
\begin{align}
	h&\left(t\right)=\nonumber\\&\sum_{l=2}^{\infty}\sum_{m=-l}^{l}{\rm e}^{-2i\varphi_{\rm L}\left(t\right)}~_{-2}Y^l_{~m}\left[-\theta_{\rm L}\left(t\right),-\psi_{\rm L}\left(t\right)\right]h^{\rm QA}_{lm}\left(t\right),\label{eq:handhqa}
\end{align}
	where $~_{-2}Y^l_{~m}$ is the spin-weighted spherical harmonics, $h^{\rm QA}_{lm}\left(t\right)$ is the $(l,m)$ mode in the co-precessing frame, and $\psi_{\rm L}\left(t\right)$ is the angle defined by
\begin{align}
	\psi_{\rm L}\left(t\right)=-\int^t_0 {\dot \varphi_{\rm L}\left(t'\right)}\cos\theta_{\rm L}\left(t'\right)dt',\label{eq:def_psi}
\end{align}
 which comes from the minimal rotation condition of co-precessing frame~\citep{2011PhRvD..84l4011B}. The initial value of $ \psi_{\rm L}$ can be chosen arbitrarily, and we set it to be zero in this work. Here, we assume that the time scales of the orbital precession and the gravitational-radiation reaction are much longer than the orbital period. Then, as the waveforms in the co-precessing frame have similar features to non-precessing waveforms, we can approximately decompose $h^{\rm QA}_{lm}$ into slowly evolving amplitude part, $A^{\rm QA}_{lm}\left(t\right)$ and rapidly evolving phase part, ${\rm e}^{-im\Phi^{\rm QA}\left(t\right)}$. This has been verified with the post-Newtonian waveforms in Ref.~\citep{2009PhRvD..79j4023A}. Here, $\Phi^{\rm QA}\left(t\right)$ is the orbital phase of the binary defined by the half of the phase of $(l,m)=(2,2)$ mode in the co-precessing frame. We note that $\Phi^{\rm QA}\left(t\right)$ is slightly different from the orbital phase in the standard post-Newtonian framework which is defined with respect to the relative coordinate separation of the binary (see Ref.~\citep{2009PhRvD..79j4023A}). Then, we can rewrite Eq.~\eqref{eq:handhqa} as
\begin{align}
	h\left(t\right)
	\approx \sum_{l=2}^{\infty}\sum_{m=-l}^{l}{\rm e}^{-2i\varphi_{\rm L}\left(t\right)}~_{-2}Y^l_{~m}\left[-\theta_{\rm L}\left(t\right),0\right]A^{\rm QA}_{lm}\left(t\right)\nonumber\\
	\times{\rm e}^{-im\left[\Phi^{\rm QA}\left(t\right)+\psi_{\rm L}\left(t\right)\right]}.\label{eq:hinaqa}
\end{align}
	Equation~\eqref{eq:hinaqa} shows that the waveforms in the inertial frame can be described by the superposition of the wave components for which the phase is $-m \Phi\left(t\right)=-m\left[\Phi^{\rm QA}\left(t\right)+\psi_{\rm L}\left(t\right)\right]$, with relatively slowly evolving part of ${\rm e}^{-2i\varphi_{\rm L}\left(t\right)}~_{-2}Y^l_{~m}\left[-\theta_{\rm L}\left(t\right),0\right]A^{\rm QA}_{lm}\left(t\right)$. In particular, the dominant modes of gravitational waves are contained in the wave components with $(l,m)=(2,\pm 2)$.
	
	If  $\Phi \left(t\right)$ is known a priori, we can decompose each wave component in Eq.~\eqref{eq:hinaqa} by performing a transformation as
\begin{align}
	{\tilde h}\left(m\right)=\int^{\infty}_{-\infty}  h\left(t\right) {\rm e}^{-im\Phi\left(t\right)}{\dot \Phi}\left(t\right)dt.\label{eq:md}
\end{align}
	This transformation, which we refer to as the {\it mode decomposition} in the following, is the Fourier transformation of $h$ not with respect to time but with respect to the phase $\Phi$. We can easily reconstruct the time-domain waveforms from the mode spectrum, ${\tilde h}\left(m\right)$, by the inverse transformation,
\begin{align}
	h\left(t\right)=\frac{1}{2\pi}\int^{\infty}_{-\infty}{\tilde h}\left(m\right){\rm e}^{im\Phi\left(t\right)}dm.\label{eq:imd}
\end{align}

	We note that there is practically a degree of freedom in the choice of the phase variable for the mode decomposition. For example, if we consider $\alpha\left(t\right)$ as a function which evolves much slower than $\Phi\left(t\right)$, and employ $\Phi+\alpha$ as a phase variable for the mode decomposition, Eq.~\eqref{eq:md} leads to
\begin{align}
	{\tilde h}\left(m\right)&=\int^{\infty}_{-\infty}  h\left(t\right) {\rm e}^{-im\left[\Phi\left(t\right)+\alpha\left(t\right)\right]}\left[{\dot \Phi}\left(t\right)+{\dot \alpha}\left(t\right)\right]dt\nonumber\\
	&\approx\int^{\infty}_{-\infty}  h\left(\Phi\right) {\rm e}^{-im\left[\Phi+\alpha\left(\Phi\right)\right]}d\Phi\nonumber\\
	&\approx {\rm e}^{-im\alpha_0}\int^{\infty}_{-\infty}  h\left(\Phi\right) {\rm e}^{-im\left(1+\alpha'_0\right)\Phi}d\Phi\nonumber\\	
	&={\rm e}^{-im\alpha_0}{\tilde h}\left[m\left(1+\alpha'_0\right)\right].\label{eq:md2}
\end{align}
Here, $\alpha_0$ and $\alpha'_0$ denote the value of $\alpha$ and $d\alpha/d\Phi(\ll1)$ at $\Phi=0$, respectively. For the transformation from the second line to the third line, we expand the $\alpha$ up to the linear order of $\Phi$ and neglect the higher order terms because the time evolution of $\alpha$ is much smaller than that of $\Phi$. Equation~\eqref{eq:md2} shows that the mode spectrum is shifted only slightly, and its amplitude does not change by the change of the phase variable. Therefore, adding a slowly evolving function, for example $\varphi_{\rm L}$, to the phase variable has only a minor effect on the mode decomposition and, in particular, on the extraction of the dominant modes, of which procedure is introduced in Sec.~\ref{sec:2_c}.


\begin{figure}
	\begin{center}
		\includegraphics[width=90mm]{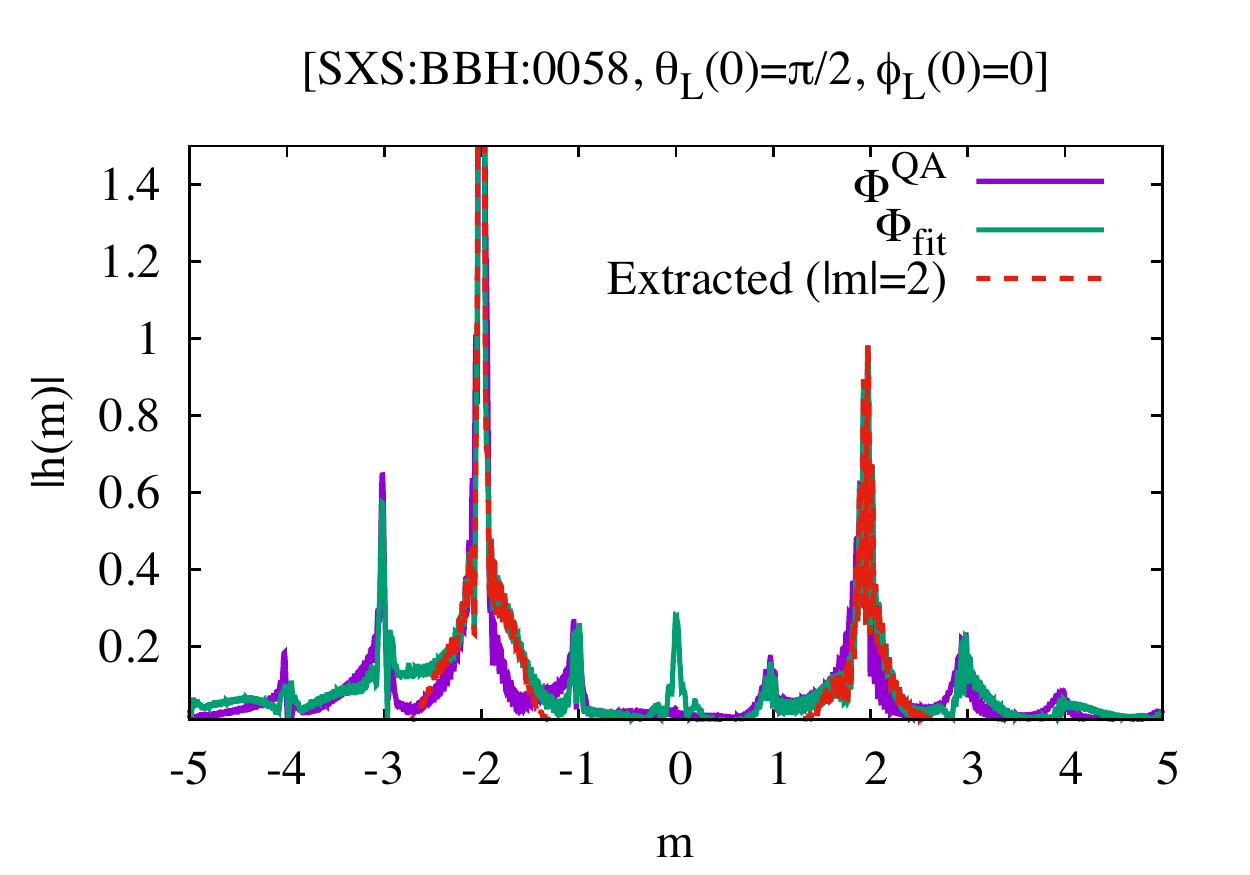}
	\end{center}
	\caption{The mode spectra of the gravitational waveforms from a precessing binary (SXS:BBH:0058). We used the waveforms observed from the direction in which $\theta_{\rm L}=\pi/2$ and $\varphi_{\rm L}=0$ are satisfied at the initial time of the simulation. The curves ``$\Phi^{\rm QA}$'' and ``$\Phi_{\rm fit}$'' show the mode spectra using $\Phi^{\rm QA}$ and $\Phi_{\rm fit}$ as the orbital phase in Eq.~\eqref{eq:md}, respectively. The curve ``Extracted $(|m|=2)$'' shows the mode spectra to which a window function Eq.~\eqref{eq:mwindow} is applied.}
	\label{fig:mode_spec}
\end{figure}	
	
	As an illustration, we perform the mode decomposition of precessing waveforms using the waveforms derived in numerical relativity simulations. As an example, we employ SXS:BBH:0058 in Refs.~\citep{SXS:catalog,Mroue:2012kv,Mroue:2013xna}, which is a waveform of a binary black hole for the case that the mass ratio is $5$, only the larger mass black hole has a dimensionless spin with $0.5$, and the black-hole spin initially lies in the orbital plane. In this model, the orbital angular momentum is misaligned with the initial total angular momentum by $\approx0.5\,{\rm rad}$, and $\approx1$ cycle of the precession occurs before the merger. We note that this precession range is similar to the one of the model employed in Ref.~\citep{2012PhRvD..85h4003O}, where the complexity of the precessing waveforms is discussed. We generate the complex waveform strain observed from a specific direction by employing all the components of spherical harmonics up to $l=8$ in the inertial frame. In this section, we specifically choose the direction of the observation that satisfies $\theta_{\rm L}=\pi/2$ and $\varphi_{\rm L}=0$ at the initial time of the simulation. The results for different directions are shown in Sec.~\ref{sec:res}. We use  $\Phi^{\rm QA}\left(t\right)$ as the phase variable for performing the mode decomposition in Eq.~\eqref{eq:md}.
	
	Figure~\ref{fig:mode_spec} plots the resulting mode spectrum (see the plot referred to as “$\Phi^{\rm QA}$”). This shows that the mode spectrum has peaks at integer values of $m$, and each peak is clearly separated. This suggests that we can approximately extract $m$-mode wave components of $h$ by performing the mode decomposition Eq.~\eqref{eq:md}. Then, applying an appropriate filter or window function to the mode spectrum ${\tilde h}\left(m\right)$, it is possible to approximately reconstruct the time-domain waveforms by Eq.~\eqref{eq:imd} (see Sec.~\ref{sec:2_c}). 

\subsection{Extracting the orbital phase}\label{sec:ext}

\begin{figure}
	\begin{center}
		\includegraphics[width=90mm]{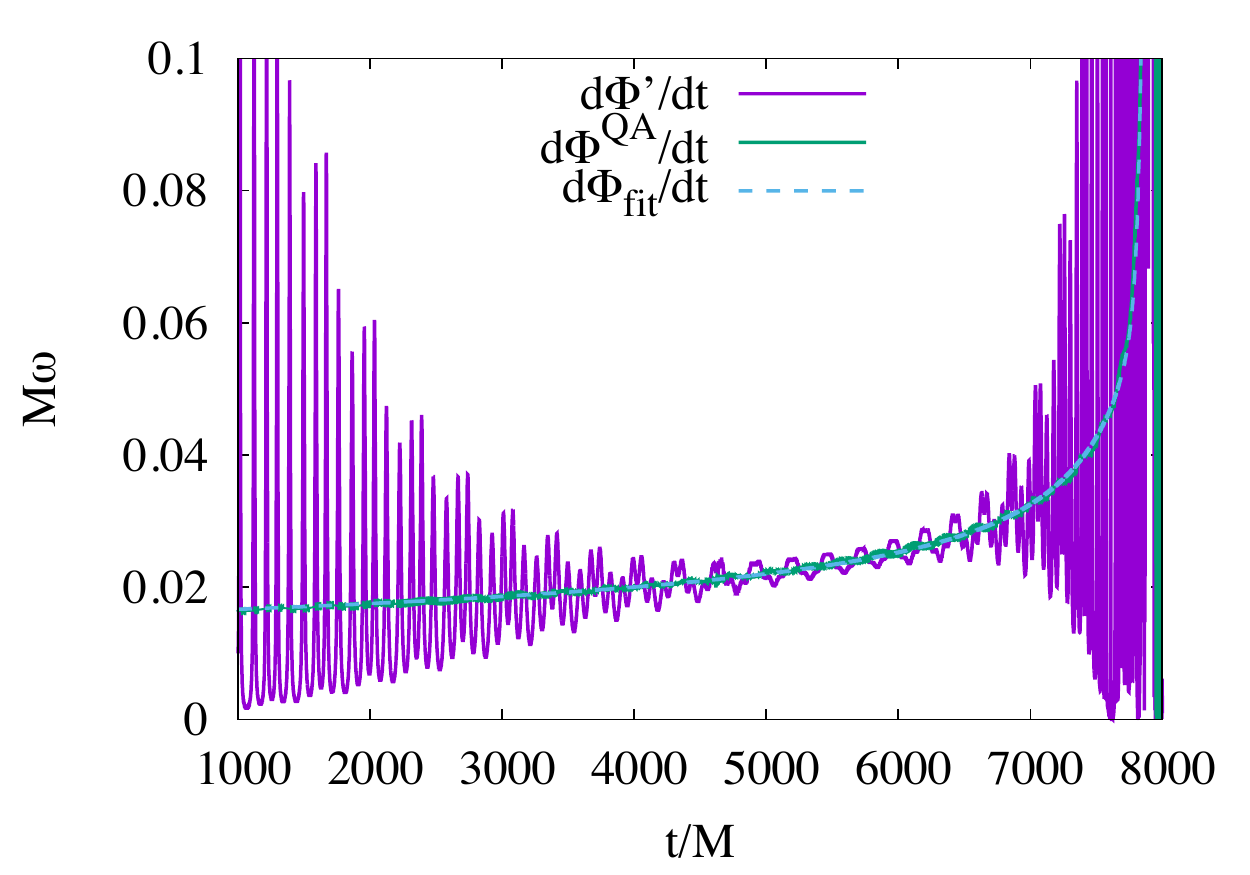}
	\end{center}
	\caption{The comparison of orbital frequencies obtained by several methods. The curves ``$d\Phi'/dt$'', ``$d\Phi^{\rm QA}/dt$'', and ``$d\Phi_{\rm fit}/dt$'' show orbital frequencies obtained by the time derivative of $\Phi'$, $\Phi^{\rm QA}$, and $\Phi_{\rm fit}$, respectively.}
	\label{fig:comp_ome}
\end{figure}

	To practically perform the mode decomposition of the observed waveforms themselves, we need to extract the orbital phase of the binary from the observational data. If the binary is not precessing and the observer is located along the orbital axis, the orbital phase of the binary in the inspiral orbits can be extracted from the waveforms by
\begin{align}
	\Phi'\left(t\right)=\frac{1}{2}\int^t_0 \frac{\left|{\rm Im}\left[h^*\left(t'\right) {\dot h\left(t'\right)}\right]\right|}{|h\left(t'\right)|^2}dt',\label{eq:phi_orb}
\end{align}
where $h^*$ denotes the complex conjugate of $h$. However, if the binary is precessing, we cannot obtain the orbital phase directly from Eq.~\eqref{eq:phi_orb}. In Fig.~\ref{fig:comp_ome}, we plot the time derivative of $\Phi'\left(t\right)$ calculated by Eq.~\eqref{eq:phi_orb} as well as that of $\Phi^{\rm QA}$ for the same waveforms as those used in Fig.~\ref{fig:mode_spec}. We find that ${\dot \Phi}'\left(t\right)$ is strongly oscillating due to the mixing of wave components with different frequencies, while ${\dot \Phi}^{\rm QA}\left(t\right)$ evolves monotonically. Therefore, we cannot use $\Phi'\left(t\right)$ directly for the mode decomposition.

	Instead of employing $\Phi'\left(t\right)$, we have to extract the non-oscillatory part of $\Phi'\left(t\right)$ for the mode decomposition. While $\Phi'\left(t\right)$ oscillates strongly, it still behaves in a similar manner to $\Phi^{\rm QA}\left(t\right)$ if we take the time average. Therefore, we expect that we can approximately extract the ``orbital phase'' which can be used for the mode decomposition if we remove the oscillation from $\Phi'\left(t\right)$. In this work, we extract the non-oscillatory part by fitting $\Phi'\left(t\right)$ with a non-oscillating function defined by
\begin{align}
	\Phi_{\rm fit}\left(t\right)=\left\{
	\begin{array}{cc}
		\Phi_{\rm fit}^{\rm ins}\left(t\right) & t\le t_0,\\
		b_0+c_1\left(t-t_0\right)+b_1\left[{\rm e}^{-\left(t-t_0\right)/c_2^2}-1\right] & t\ge t_0,
	\end{array}
	\right.\label{eq:fit}
\end{align}
where $\Phi_{\rm fit}^{\rm ins}\left(t\right)$ is the inspiral part defined by
\begin{align}
	\Phi_{\rm fit}^{\rm ins}\left(t\right)&=a+a_1\left(t_1-t\right)^{5/8}+a_2\left(t_1-t\right)^{3/8}\nonumber\\
	&+a_3\left(t_1-t\right)^{1/4}+a_4\left(t_1-t\right)^{1/8}+a_5^{(1)}{\rm ln}\left(t_1-t\right)\nonumber\\
	&+\left[a_6^{(0)}+a_6^{(1)}{\rm ln}\left(t_1-t\right)\right]\left(t_1-t\right)^{-1/8}\nonumber\\
	&+a_7\left(t_1-t\right)^{-1/4}.\label{eq:fit_ins}
\end{align}
Here, the functional form of Eq.~\eqref{eq:fit_ins} is motivated by Taylor-T3 approximant~\citep{2014LRR....17....2B,2009PhRvD..80h4043B}. For this prescription, $t_0$ is taken to be the time of global maximum of $|h\left(t\right)|$ ($t_0=7857\,M$ for this case), and $b_0$ and $b_1$ are chosen so that $\Phi_{\rm fit}\left(t\right)$ and ${\dot \Phi}_{\rm fit}\left(t\right)$ are continuous at $t=t_0$. We determine $a_n~\left(n=1,\cdots,7\right)$, $c_1$, $c_2$ and $t_1>t_0$ by the least-square fitting method using $\Phi'\left(t\right)$. We only use the data from $t=1000\,M$ ($t=0$ is the time at which the simulation starts) to the time that $|h\left(t\right)|$ becomes smaller than $5\%$ of its peak value for the first time. This time window is 
chosen to avoid the unphysical modulation in the beginning of the simulation and the unimportant part of the waveforms after the onset of merger. 

	In Fig.~\ref{fig:comp_ome}, we plot ${\dot \Phi}_{\rm fit}\left(t\right)$. 
We find that $\dot \Phi_{\rm fit}$ agrees with ${\dot \Phi}^{\rm QA}\left(t\right)$ within $2\%$ up to the merger. We also plot the mode spectrum of the waveform obtained using $\Phi_{\rm fit}\left(t\right)$ in Eq.~\eqref{eq:md} in Fig.~\ref{fig:mode_spec}. Although there is slight deviation from the ones obtained using $\Phi^{\rm QA}\left(t\right)$, we find that the mode spectrum has peaks in integer values of $m$ and each peak is clearly separated. This suggests that $\Phi_{\rm fit}\left(t\right)$ can be a good substitute for $\Phi^{\rm QA}\left(t\right)$ to perform the mode decomposition. We note that ${\dot \Phi}_{\rm fit}\left(t\right)$ does not strictly agree with neither ${\dot \Phi}\left(t\right)$ nor ${\dot \Phi}^{\rm QA}\left(t\right)$, but rather agrees well with ${\dot \Phi}\left(t\right)+{\dot \varphi}_{\rm L}\left(t\right)~{\rm sign}\left[\cos \theta_{\rm L}\left(t\right)\right]$. We can also prove this analytically by assuming that $(l,m)=(2,\pm2)$ modes in the co-precessing frame are the dominant modes. Because $|{\dot \varphi}_{\rm L}|$ is much smaller than $|{\dot \Phi}\left(t\right)|$, the deviation of $\Phi_{\rm fit}\left(t\right)$ from $\Phi\left(t\right)$ only weakly affect the accuracy of the mode decomposition at least for extracting the dominant modes in the inspiral orbits.

\subsection{Extracting the wave components}\label{sec:2_c}
	We introduce here a window function to extract specific wave components in the mode spectra. We define a one-sided amplitude of the mode spectra by
\begin{align}
	A\left(m\right):=\sqrt{|{\tilde h}\left(m\right)|^2+|{\tilde h}\left(-m\right)|^2}.\label{eq:hmamp}
\end{align}
$A\left(m\right)$ has the largest peak in $|m|\approx 2$, and small side peaks in integer values of $m$. As we mentioned above, the information of the dominant modes of gravitational waves is contained primarily in the modes of $|m|=2$. To single out only the information around $|m|\approx 2$, we performed the extraction in following three steps. First, we fit $A\left(m\right)$ around $|m|\approx 2$ by a Lorentzian function, 
\begin{align}
	L\left(m;A_0,m_{\rm p},m_{1/2}\right)=\frac{A_0}{1+(m-m_{\rm p})^2/ m_{1/2}^2},
\end{align}
where $A_0$ is the peak amplitude, $m_{\rm p}$ and $m_{1/2}$ are the location of the peak and the half-width at half maximum, respectively. We perform the least-square fitting to determine these fitting parameters. We note that $m_{\rm p}$ is also a fitting parameter, while its initial guess is set to be $2$. We find that the value after the fitting deviates from the initial value only by $\approx0.02$. 

Next, we introduce a window function $w\left(m\right)$ defined by
\begin{align}
	&w\left(m\right)=\nonumber\\
	&\left\{
	\begin{array}{cc}
		1 & |m-m_{\rm p}|<\Delta m_1,\\\\
		\displaystyle H\left(m\right)+\left[1-H\left(m\right)\right]\frac{L\left(m\right)}{A\left(m\right)} & \Delta m_1\le |m-m_{\rm p}|<\Delta m_2,\\
		\displaystyle \frac{L\left(m\right)}{A\left(m\right)} &  \Delta m_2\le|m-m_{\rm p}|,
	\end{array}
	\right.\label{eq:mwindow}
\end{align}
where
\begin{align}
	 H&\left(m\right)=\frac{1}{2}\left[1+\cos\left(\pi\frac{|m-m_{\rm p}|-\Delta m_1}{\Delta m_2-\Delta m_1}\right)\right].
\end{align}
Here, we chose $\Delta m_1=0.35$ and $\Delta m_2=0.75$. Finally, we define the extracted mode spectrum ${\tilde h}^{\rm ext}\left(m\right)$ by ${\tilde h}^{\rm ext}\left(m\right)=w\left(m\right){\tilde h}\left(m\right)$. 

\begin{figure}
	\begin{center}
		\includegraphics[width=90mm]{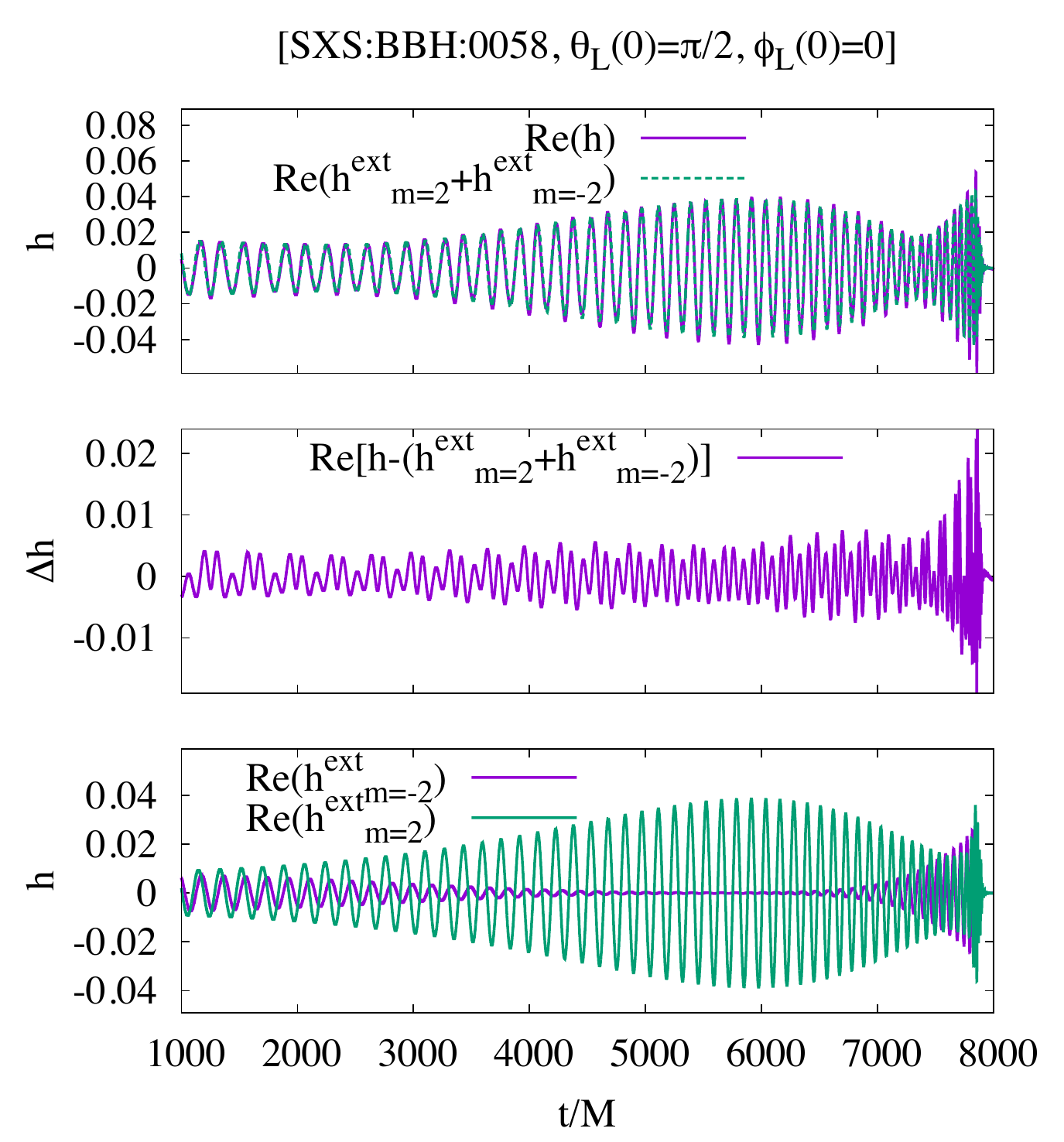}
	\end{center}
	\caption{The comparison of the original and extracted waveforms in the time domain. The upper panel shows the real part of the original complex waveform strain, $h$, and sum of $m=\pm2$ wave components extracted from the mode spectrum with respect to $\Phi_{\rm fit}\left(t\right)$, and the middle panel shows the difference between those two waveforms. The bottom panel shows the real part of $m=2$ and $m=-2$ mode wave components.}
	\label{fig:h_ext_comp}
\end{figure}

	Applying this window function,  the amplitude of the mode spectra in $|m-m_{\rm p}|>\Delta m_1$ is continuously suppressed and normalized to $L\left(m\right)$, and the peaks in $|m|\neq2$  are suppressed. Indeed, a plot for the extracted mode spectrum, ${\tilde h}^{\rm ext}\left(m\right)$, in Fig.~\ref{fig:mode_spec} shows that only the peaks in $|m|=2$ are remaining. We can then obtain the wave components for $m=2$ and $-2$ in the time domain, $h^{\rm ext}_{m=2}\left(t\right)$ and $h^{\rm ext}_{m=-2}\left(t\right)$, by performing the inverse transformation of the spectra for each peak using $\Phi_{\rm fit}\left(t\right)$. Figure~\ref{fig:h_ext_comp}  compares the original and extracted waveforms in the time domain. The original waveforms and sum of $m=\pm2$ wave components agree approximately with each other (see the upper panel in Fig.~\ref{fig:h_ext_comp}), and the difference between these two waveforms oscillates in different frequency from the dominant-mode frequency (see the middle panel in Fig.~\ref{fig:h_ext_comp}). This suggests that wave components of $m\neq\pm2$ are removed and only dominant wave components are extracted from the original strain. The bottom panel in Fig.~\ref{fig:h_ext_comp} shows the real part of $m=2$ and $m=-2$ mode wave components. The smooth change in the amplitude reflects the orbital precession (see Eq.~\eqref{eq:ext_m2}).

\subsection{Extracting the instantaneous orbital axis}

	Assuming that the extracted wave components of $|m|=2$ are dominated by the $l=2$ components of the spherical harmonics, Eq.~\eqref{eq:hinaqa} gives the description for these wave components as
\begin{align}
h^{\rm ext}_{m=\pm2}\left(t\right)\approx\frac{1}{8}\sqrt{\frac{5}{\pi}}\left[1\pm\cos \theta_{\rm L}\left(t\right)\right]^2A^{\rm QA}_{22}\left(t\right){\rm e}^{-2i\left[\varphi_{\rm L}\left(t\right)\pm\Phi\left(t\right)\right]}.\label{eq:ext_m2}
\end{align}

	If we further assume that the system has an approximate equatorial symmetry in the co-precessing frame \footnote{Strictly speaking, this is not true as Ref.~\citep{2014arXiv1409.4431B} has pointed out that there remains some asymmetric modulation in the waveforms even in the co-precessing frame. However, while it can be the source of error in the analysis, we neglect such a contribution in this paper since it is expected to be small.}, and hence, $A^{\rm QA}_{22}\left(t\right)=A^{\rm QA}_{2-2}\left(t\right)$ holds, we can measure $\theta_{\rm L}\left(t\right)$, $\varphi_{\rm L}\left(t\right)$, and $\Phi\left(t\right)$ by
\begin{align}
	\theta_{\rm L}\left(t\right)=\cos^{-1}\left[\frac{\sqrt{\left|h^{\rm ext}_{m=2}\left(t\right)\right|}-\sqrt{\left|h^{\rm ext}_{m=-2}\left(t\right)\right|}}{\sqrt{\left|h^{\rm ext}_{m=2}\left(t\right)\right|}+\sqrt{\left|h^{\rm ext}_{m=-2}\left(t\right)\right|}}\right],\label{eq:ext_th}
\end{align}
\begin{align}
	{\rm e}^{-4i\varphi_{\rm L}\left(t\right)}=\frac{h^{\rm ext}_{m=2}\left(t\right)h^{\rm ext}_{m=-2}\left(t\right)}{\left|h^{\rm ext}_{m=2}\left(t\right)\right|\left|h^{\rm ext}_{m=-2}\left(t\right)\right|},\label{eq:ext_varphi}
\end{align}
and
\begin{align}
	{\rm e}^{-4i\Phi\left(t\right)}=\frac{h^{\rm ext}_{m=2}\left(t\right)h^{\rm ext,*}_{m=-2}\left(t\right)}{\left|h^{\rm ext}_{m=2}\left(t\right)\right|\left|h^{\rm ext}_{m=-2}\left(t\right)\right|}.\label{eq:ext_phi}
\end{align}
$\Phi^{\rm QA}\left(t\right)$ is determined from $\Phi\left(t\right)$ and $\psi_{\rm L}\left(t\right)$, where $\psi_{\rm L}\left(t\right)$ is determined by Eq.~\eqref{eq:def_psi} using the extracted result of $\theta_{\rm L}\left(t\right)$ and $\varphi_{\rm L}\left(t\right)$ (note that $\varphi_{\rm L}\left(t\right)$ is only determined up to multiple times $\pi/2$ in our method). Using $\theta_{\rm L}\left(t\right)$, we can determine $A^{\rm QA}_{22}\left(t\right)$ (or $A^{\rm QA}_{2-2}\left(t\right)$) from $\sqrt{\left|h^{\rm ext}_{m=2}\left(t\right)\right|}$ (or $\sqrt{\left|h^{\rm ext}_{m=-2}\left(t\right)\right|}$). Then, the $(l,m)=(2,\pm2)$ modes in the co-precessing frame are reconstructed by $A^{\rm QA}_{2\pm2}\left(t\right){\rm e}^{\mp2i\Phi^{\rm QA}\left(t\right)}$.

\section{Application}\label{sec:res}
	
\begin{figure}
	\begin{center}
		\includegraphics[width=90mm]{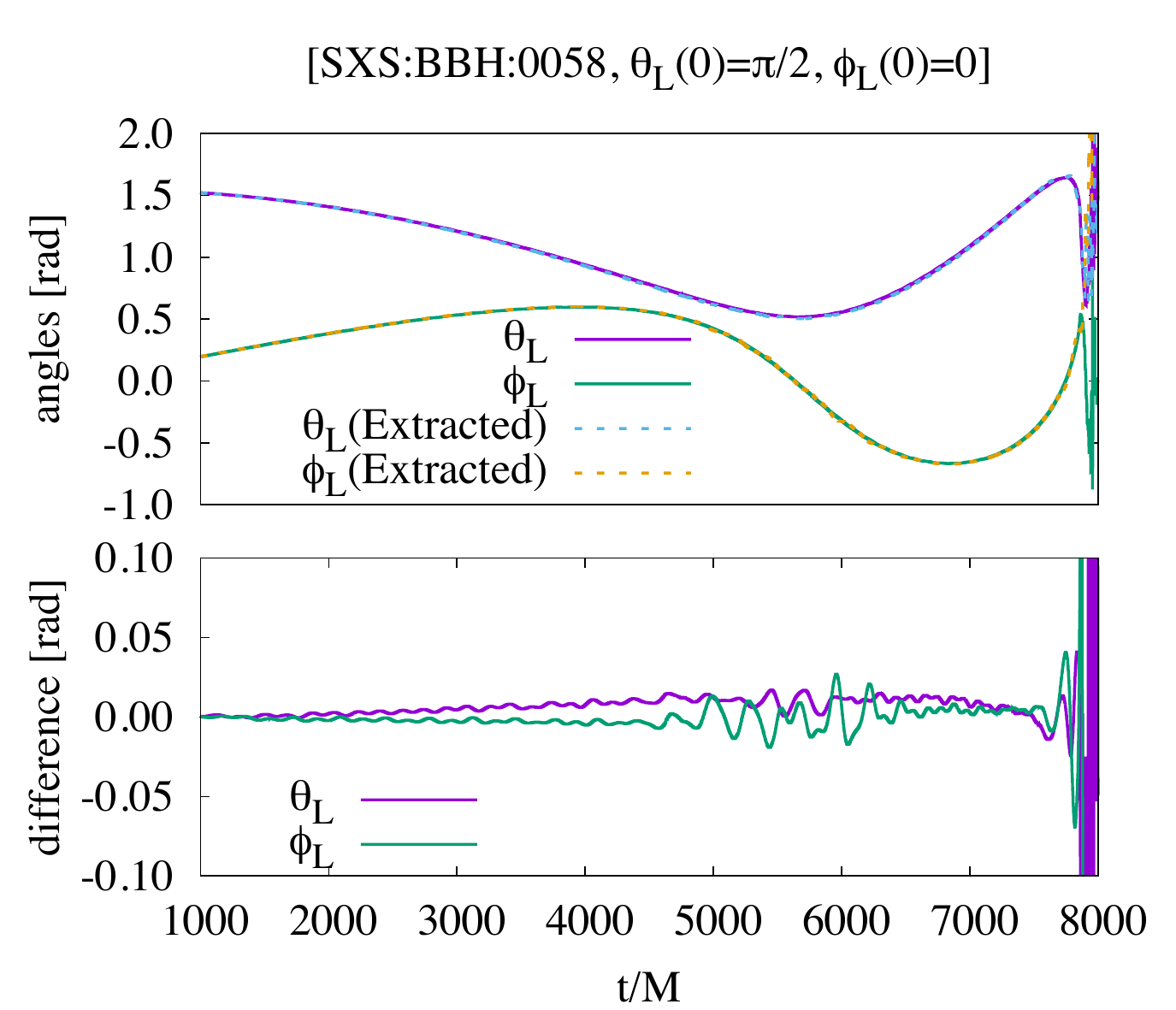}
	\end{center}
	\caption{The comparison of $\theta_{\rm L}$ and $\varphi_{\rm L}$ obtained from the original QA method and the ones obtained from the extraction procedure using the mode decomposition.}
	\label{fig:ext_angle}
\end{figure}
	In this section, we examine the extraction method introduced in the previous section. First, we use the waveform strain generated from the data of SXS:BBH:0058, which were also used in Figs.~\ref{fig:mode_spec} and~\ref{fig:comp_ome}. Figure~\ref{fig:ext_angle} shows the comparisons of $\theta_{\rm L}\left(t\right)$ and $\varphi_{\rm L}\left(t\right)$ obtained from the original QA method and the ones obtained from the extraction procedure using the mode decomposition. In this figure, we find that $\theta_{\rm L}\left(t\right)$ and $\varphi_{\rm L}\left(t\right)$ agree well between two methods, and we find the deviations are always smaller than $0.07~{\rm rad}$ until $t\approx7857\,M$.
	
	Next, we check how accurately the co-precessing waveforms are reconstructed. We compare the $(l,m)=(2,2)$ mode in the co-precessing frame obtained by the original QA method and the ones reconstructed by our method. Here, instead of using the QA waveforms directly, we take the average of the $(l,m)=(2,\pm2)$ modes, namely, ${\bar h}_{22}^{\rm QA}=\left(h_{22}^{\rm QA}+h_{2-2}^{\rm QA,*}\right)/2$ for the QA method. This average is taken so that the equatorial symmetry in the co-precessing frame is imposed. This is consistent with the assumption which we made in the extraction procedure. In addition, this removes the most parts of the residual modulations in $h_{22}^{\rm QA}$ and $h_{2-2}^{\rm QA}$ which remain even after transforming to the co-precessing frame~\citep{2014arXiv1409.4431B,2017arXiv170100550B}. Since these modulations, for which the oscillation frequency is different from the dominant mode, are expected to be removed by the extraction procedure, it is reasonable to use the averaged waveforms. 
	
\begin{figure}
	\begin{center}
		\includegraphics[width=90mm]{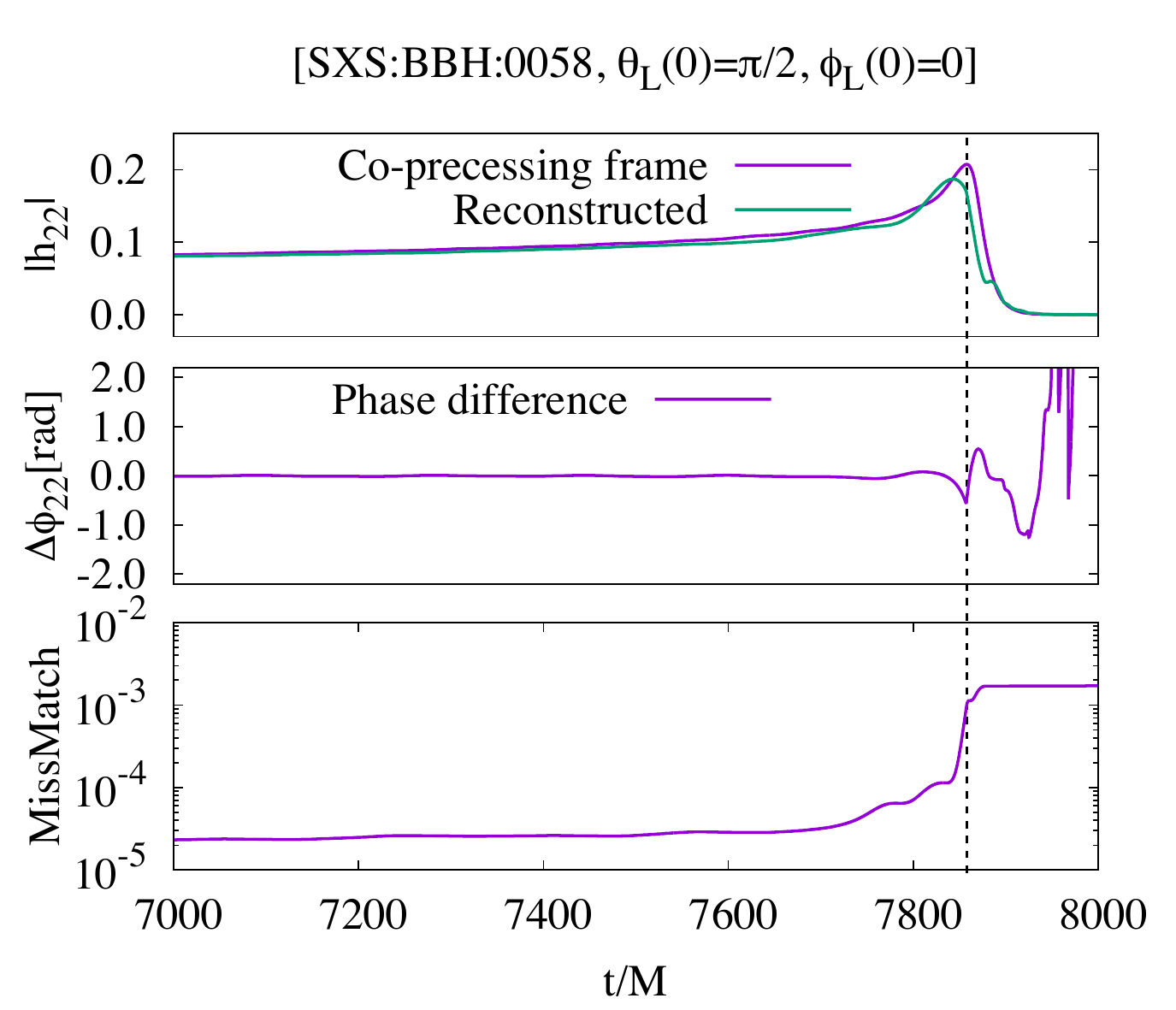}
	\end{center}
	\caption{The comparison of the co-precessing frame waveforms obtained by the QA method and the ones reconstructed from extracted waveforms using the mode decomposition. The top panel compares of the co-precessing frame amplitude of the $(l,m)=(2,2)$ mode. The middle panel shows the phase difference between two waveforms for the case that the mismatch for $(t_{\rm i},t_{\rm f})=(1000\,M,\infty)$ is the minimum (see Eq.~\eqref{eq:def_mis}). The bottom panel shows the  mismatch between the two waveforms as a function of the upper bound of integral, $t_{\rm f}$. The lower band of the integral, $t_{\rm i}$, is always set to be $1000\,M$. We note that we take the average of the $(l,m)=(2,\pm2)$ modes for the QA waveforms to impose the equatorial symmetry. The vertical dashed line denotes the peak time of the amplitude.}
	\label{fig:comp_prec}
\end{figure}

	In the top and middle panels in Fig.~\ref{fig:comp_prec}, we compare the co-precessing frame amplitude by the two methods and show the phase difference between two waveforms, respectively. The two waveforms agree well with each other in both amplitude and the phase until $t\approx7800\,M$. Their deviations are enhanced for $t=7800$--$8000\,M$, while the deviation in the phase remains smaller than $\approx0.5~{\rm rad}$ until the peak amplitude is reached. This late-time deviation is also found in the comparisons of $\theta_{\rm L}\left(t\right)$ and $\varphi_{\rm L}\left(t\right)$. We suspect that these deviations would be due to the fact that the precession timescale becomes short and comparable to the orbital period just before the merger. If we perform the mode decomposition focusing only on the waveforms after $t\approx7800\,M$, the width of the peaks in the spectra becomes broad and overlap with each other as two time scales become comparable. This suggests that some part of information in the merger-ringdown stages leaks to the other peaks in the spectra. The prescriptions for the phase fitting in Eq.~\eqref{eq:fit} and the window function in Eq.~\eqref{eq:mwindow} can also be the source for the errors. If this is the case, further improvement is needed for these functional forms. We leave the further investigation for the origin of their errors as the future task.
	
	To discuss the agreements of the waveforms more quantitatively, we define the mismatch between two complex waveform strains, $h_1$ and $h_2$, by 
\begin{align}
	{\cal M}\left(h_1,h_2\right)=1-\max_{\varphi_{\rm c}}\frac{\displaystyle {\rm Re}\left[\left(h_1|h_2{\rm e}^{i\varphi_{\rm c}}\right)\right]}{\sqrt{\left(h_1|h_1\right)}\sqrt{\left(h_2|h_2\right)}},\label{eq:def_mis}
\end{align}
where $\left(\cdot|\cdot\right)$ is the Hermitian inner product defined by
\begin{align}
	\left(h_1|h_2\right)=\int_{t_{\rm i}}^{t_{\rm f}} h_1^*\left(t\right)h_2\left(t\right)dt.\label{eq:def_inp}
\end{align}
	Here, $t_{\rm i}$ is the lower bound of the integral which is always set to be $1000\,M$ in this work, and $t_{\rm f}$ is the upper bound of the integral. We note that our definition of the mismatch is different from the usual one that is employed in previous data-analysis studies (see, e.g., Ref.~\citep{2017arXiv170100550B}). Our definition is identical to the case that the noise spectrum density of the detector is assumed to be white~\citep{2012PhRvD..86j4063S}. We employ the definition in Eqs.~\eqref{eq:def_mis} and~\eqref{eq:def_inp} in this paper because we can calculate the mismatch in the time domain and easily show in which part of the waveforms the error is induced. We find that mismatches calculated by Eqs.~\eqref{eq:def_mis} and~\eqref{eq:def_inp} for $(t_{\rm i},t_{\rm f})=(1000\,M,\infty)$ are similar to the values calculated by the usual definition of mismatch assuming $M=10\,M_\odot$ and using a designed noise curve of Advanced LIGO (for the zero-detuned high power configuration~\citep{aLIGOnoise}).
	
	In the bottom panel of Fig.~\ref{fig:comp_prec}, we plot the mismatch between the two waveforms as a function of the upper bound of integral, $t_{\rm f}$. We find that the mismatch is always below $10^{-4}$ until the time of peak amplitude, and rapidly increases to the order of $10^{-3}$ after the peak time. This shows that the reconstructed waveforms have the largest error around the time of peak amplitude.

\begin{table}
\begin{tabular}{c|c|c|c|c}\hline
		&							&				&	${\cal M}$		&	${\cal M}$\\
Model	&$~\theta_{\rm L}\left(0\right)~$	&	$f_{2\pm2}$	&$~t_{\rm i}=1000\,M~$	&$t_{\rm i}=1000\,M$	\\
		&							&				&$t_{\rm f}=\infty$		&$t_{\rm f}=t_0-100\,M$	\\\hline
SXS:BBH:0058		&	$0$		&	~0.020~	&	$1.13\times 10^{-3}$	&	$2.13\times 10^{-4}$\\
					&	$\pi/4$	&	0.018	&	$1.40\times 10^{-3}$	&	$1.66\times 10^{-4}$\\
$m_1/m_2=5$		&	$\pi/2$	&	0.038	&	$1.71\times 10^{-3}$	&	$5.58\times 10^{-5}$\\
${\bf S}_1=(0.5,0,0)$	&	$3\pi/4$	&	0.055	&	$2.54\times 10^{-3}$	&	$1.81\times 10^{-4}$\\
${\bf S}_2={\bf 0}$		&	$\pi$	&	0.020	&	$1.02\times 10^{-3}$	&	$1.19\times 10^{-4}$\\\hline

SXS:BBH:0037		&	$0$		&	0.007	&	$4.05\times 10^{-3}$	&	$1.11\times 10^{-3}$\\
					&	$\pi/4$	&	0.009	&	$5.93\times 10^{-3}$	&	$5.03\times 10^{-3}$\\
$m_1/m_2=3$		&	$\pi/2$	&	0.037	&	$1.32\times 10^{-3}$	&	$8.36\times 10^{-5}$\\
${\bf S}_1=(0.5,0,0)$	&	$3\pi/4$	&	0.031	&	$1.81\times 10^{-3}$	&	$2.31\times 10^{-4}$\\
${\bf S}_2={\bf 0}$		&	$\pi$	&	0.007	&	$9.18\times 10^{-3}$	&	$2.95\times 10^{-3}$\\\hline
				
SXS:BBH:0164		&	$0$		&	0.001	&	$6.94\times 10^{-3}$	&	$5.73\times 10^{-3}$\\
					&	$\pi/4$	&	0.001	&	$2.42\times 10^{-3}$	&	$5.56\times 10^{-4}$\\
$m_1/m_2=1$		&	$\pi/2$	&	0.005	&	$1.35\times 10^{-3}$	&	$8.41\times 10^{-5}$\\
${\bf S}_1={\bf S}_2$	&	$3\pi/4$	&	0.002	&	$3.17\times 10^{-4}$	&	$1.31\times 10^{-5}$\\
$=(0.52,0,0.3)$		&	$\pi$	&	0.001	&	$6.93\times 10^{-3}$	&	$5.72\times 10^{-3}$\\\hline
\end{tabular}
\caption{The mismatch between the $(l,m)=(2,2)$ mode in the co-precessing frame obtained by the original QA method and the one obtained by our method. The first left column shows the model names in SXS Catalog~\citep{SXS:catalog,Mroue:2012kv,Mroue:2013xna,Hinder:2013oqa} as well as their mass ratios and their black-hole spins. The numbers in the brackets describe the $x$, $y$, and $z$ components of the black-hole spin in the source frame with $\theta_{\rm L}=0$ and $\varphi_{\rm L}=0$. The second left column shows the initial values of $\theta_{\rm L}$, which describe the initial direction of the observer with respect to the orbital axis. Here, we chose the observer so that the initial values of $\varphi_{\rm L}$ are $0$. The third column shows the relative contribution of the dominant modes in the strain defined by Eq.~\eqref{eq:domcon}. The fourth and fifth columns show the mismatches employing $\left(t_{\rm i},t_{\rm f}\right)=\left(1000\,M,\infty\right)$ and $\left(t_{\rm i},t_{\rm f}\right)=\left(1000\,M,t_0-100\,M\right)$, respectively. We note that our definition of the mismatch is different from the usual one that is employed in previous data-analysis studies (see the sentences below Eq.~\eqref{eq:def_inp}.).}
\label{tb:mismatch}
\end{table}

	To further show the usefulness of our method for a variety of precessing binaries, we calculate the mismatches of the co-precessing $(l,m)=(2,2)$ mode between those obtained by the QA method and by the mode decomposition method, picking up three precessing binary black hole models, SXS:BBH:0058, SXS:BBH:0037, and SXS:BBH:0164 in SXS Catalog~\citep{SXS:catalog,Mroue:2012kv,Mroue:2013xna,Hinder:2013oqa}. For SXS:BBH:0037 and SXS:BBH:0164, the orbital angular momenta are misaligned with the initial total angular momenta by $\approx0.3$ and $0.2\,{\rm rad}$, and the numbers of the precession cycles are $\approx1$ and $2$ before the merger, respectively. For each model, we generate five complex waveform strains observed from five different inclination angles. We again set $t_{\rm i}=1000\,M$ for computing the mismatches. We compute two mismatches for each model adopting different upper bound of the integral $t_{\rm f}$. One is computed by setting $t_{\rm f}$ to be infinity, and the other is by setting $t_{\rm f}$ to be the time earlier by $100\,M$ than the peak of amplitude. The parameters of the models, the inclination angles of the observers, and the calculated mismatches are summarized in Table~\ref{tb:mismatch}. 
	
	For every waveform strain derived from SXS:BBH:0058, mismatches are always a few times $10^{-3}$ for the case that the ringdown waveforms are included ($\left(t_{\rm i},t_{\rm f}\right)=\left(1000\,M,\infty\right)$). Mismatches decrease remarkably by an order of magnitude by excluding the waveforms in the merger and ringdown stages ($\left(t_{\rm i},t_{\rm f}\right)=\left(1000\,M,t_0-100\,M\right)$). Hence, the error of the reconstructed co-precessing frame waveforms is primarily accumulated in the merger and ringdown stages, as has already been found in Fig.~\ref{fig:comp_prec}. 
	
\begin{figure}
	\begin{center}
		\includegraphics[width=90mm]{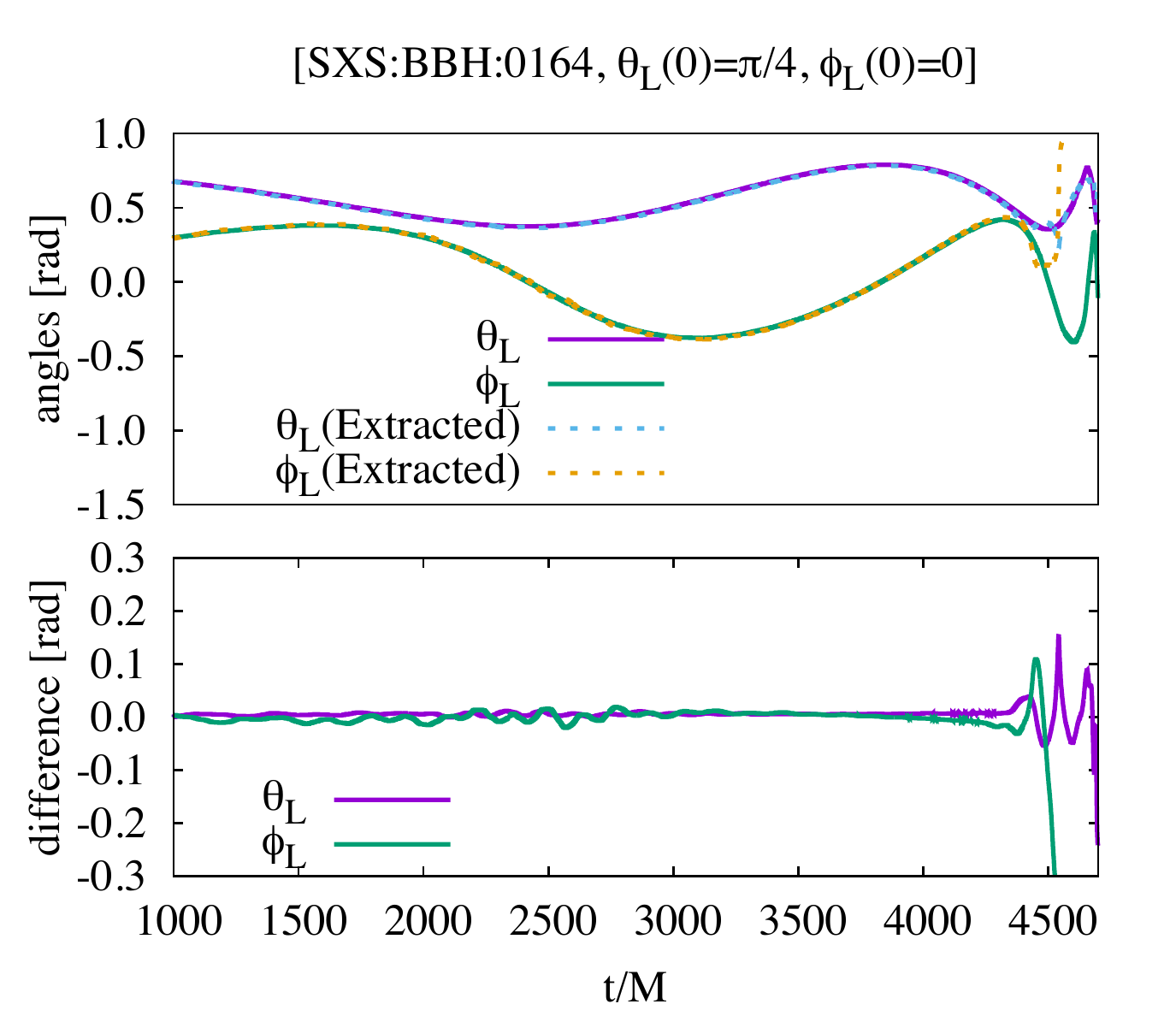}\\
		\includegraphics[width=90mm]{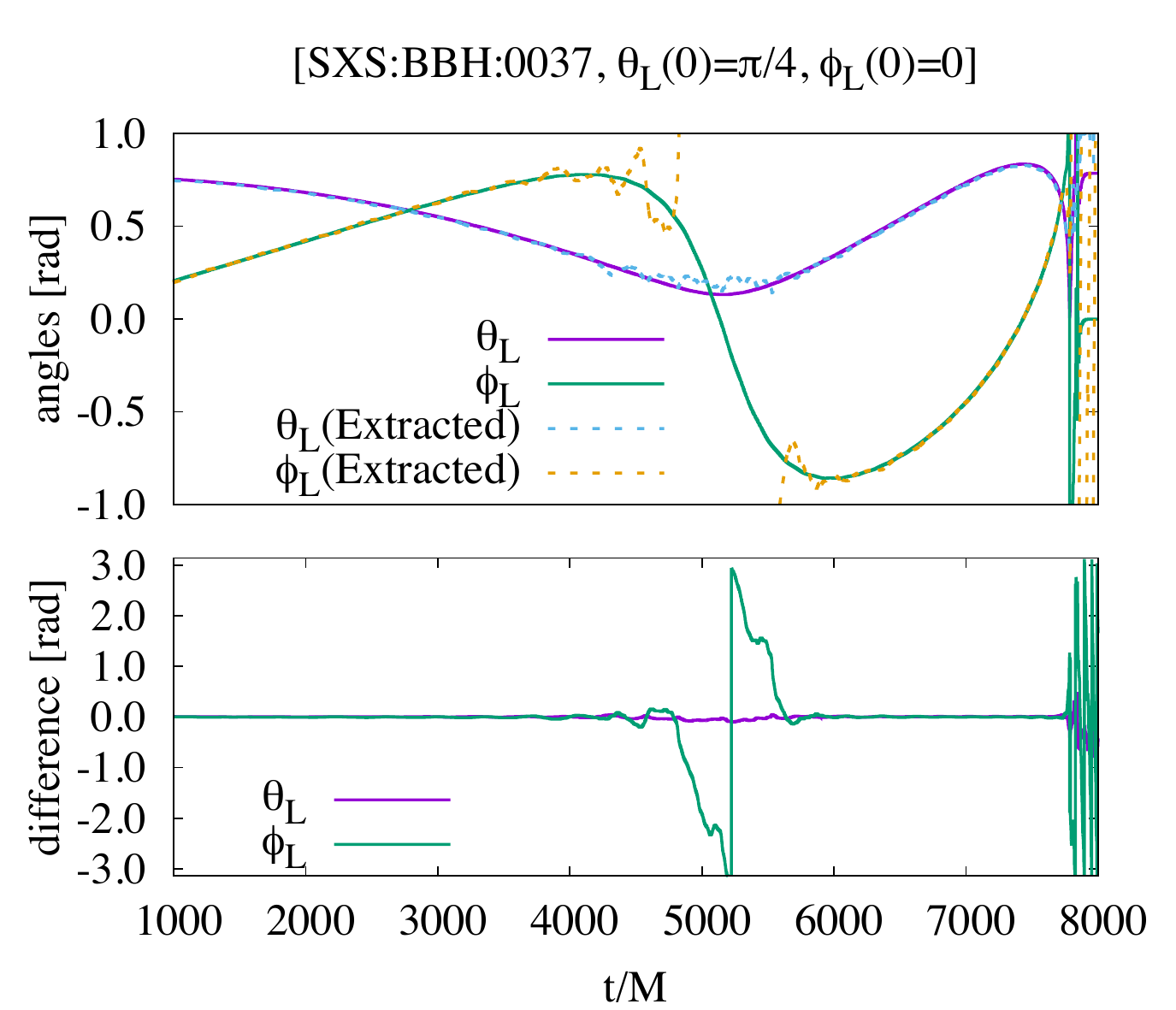}
	\end{center}
	\caption{The same as Fig.~\ref{fig:ext_angle} but for $\theta_{\rm L}\left(0\right)=\pi/4$ of SXS:BBH:0164 (top panel) and SXS:BBH:0037 (bottom panel). For the comparison, we shift the extracted result of $\varphi_{\rm L}$ by $\pi/2$ and restrict its value to $[-\pi,\pi]$ due to its uncertainty in the extraction (see Eq.~\eqref{eq:ext_varphi}).}
	\label{fig:ext_angle2}
\end{figure}

\begin{figure}
	\begin{center}
		\includegraphics[width=90mm]{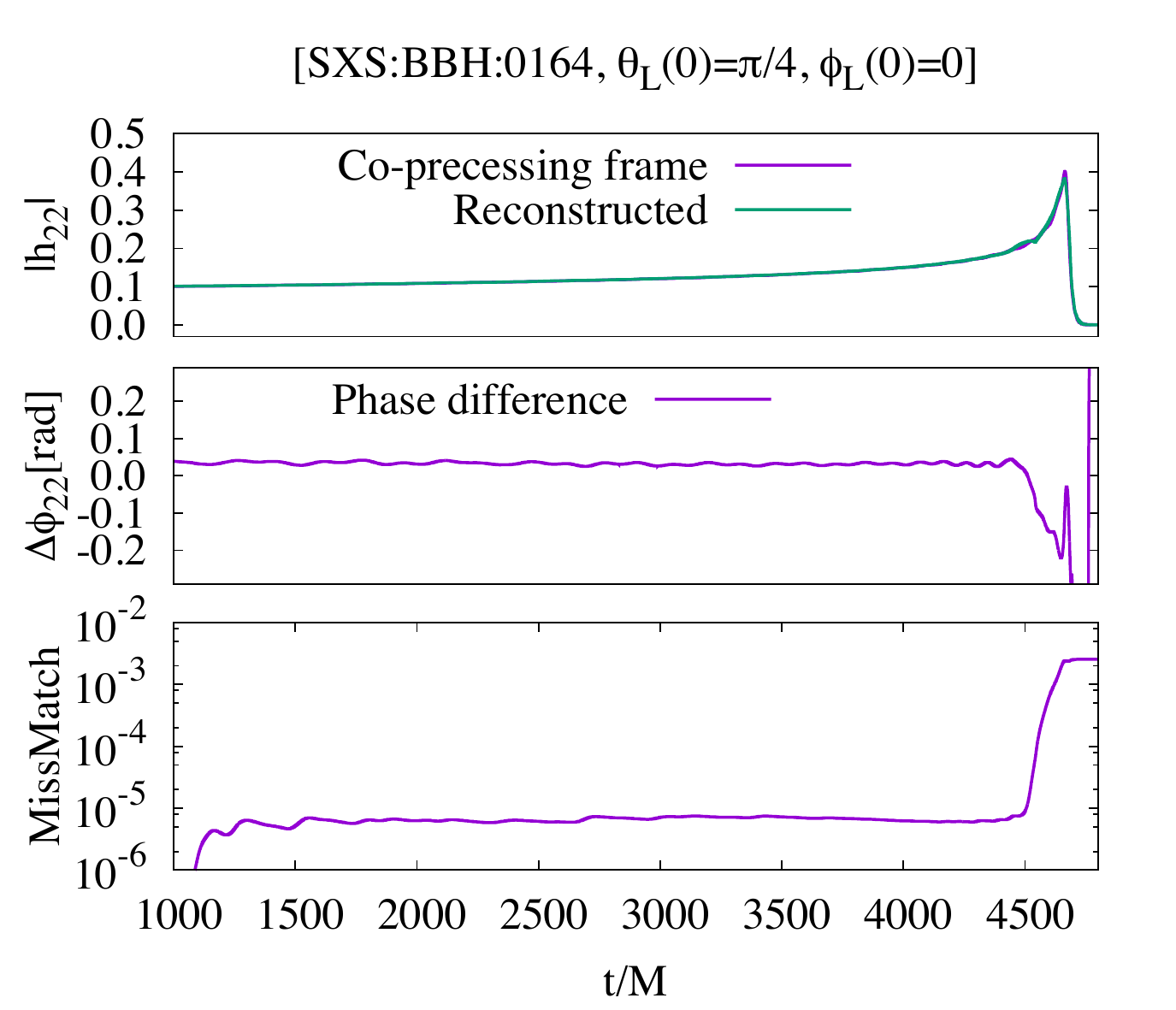}\\
		\includegraphics[width=90mm]{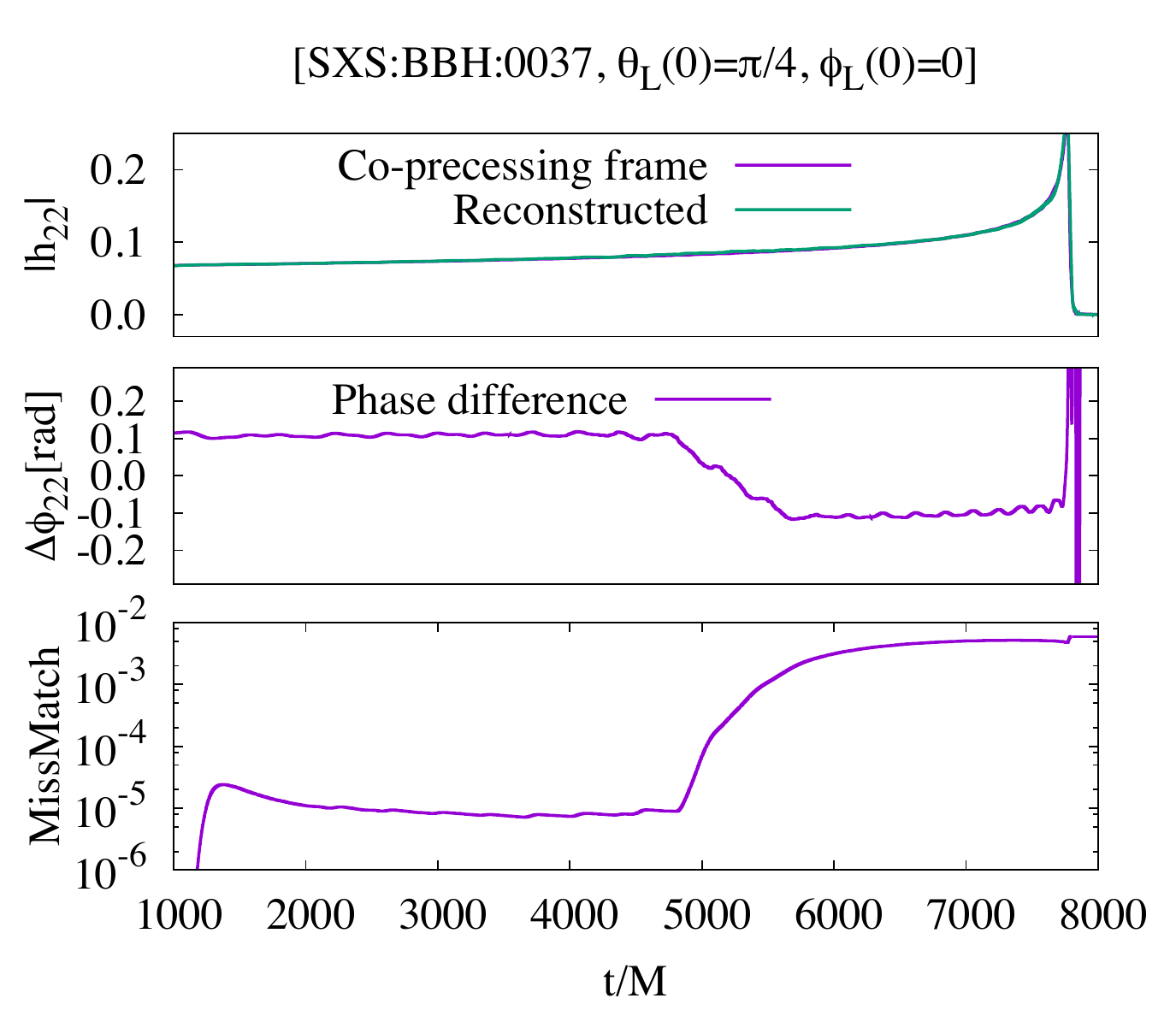}
	\end{center}
	\caption{The same as Fig.~\ref{fig:comp_prec} but for $\theta_{\rm L}\left(0\right)=\pi/4$ of SXS:BBH:0164 (top panel) and SXS:BBH:0037 (bottom panel).}
	\label{fig:comp_prec2}
\end{figure}

	For SXS:BBH:0037 and SXS:BBH:0164, the features of mismatches are quite similar to those of SXS:BBH:0058. In particular, the result for SXS:BBH:0164 shows that our extraction method can be useful not only for single-spinning binary black holes but also for double-spinning binaries (see Figs.~\ref{fig:ext_angle2} and~\ref{fig:comp_prec2}).  However, for $\theta_{\rm L}\left(0\right)=0$, $\pi/4$ and $\pi$ of SXS:BBH:0037 and $\theta_{\rm L}\left(0\right)=0$ and $\pi$ of SXS:BBH:0058, we find that the mismatches are worse than the other cases of different $\theta_{\rm L}\left(0\right)$ values or models. In particular, the improvement of mismatches by excluding the merger and ringdown stages is not as remarkable as for the other cases. This suggests that the reconstructed waveforms have errors not only in the merger and ringdown stages but also in the inspiral stage for these cases. One possible source of these errors is induced when extracting $\varphi_{\rm L}\left(t\right)$ from the waveform strain. As is seen in Eq.~\eqref{eq:ext_m2}, the norms of $h^{\rm ext}_{m=-2}$ and $h^{\rm ext}_{m=2}$ become close to zero for the case that $\theta_{\rm L}\left(t\right)$ is close to $0$ or $\pi$  (i.e., the line of sight agrees with the orbital axis), respectively. In such cases, the extraction of $\varphi_{\rm L}\left(t\right)$ becomes quite sensitive to the error in $h^{\rm ext}_{m=\pm2}$. Indeed, it is shown in Fig.~\ref{fig:ext_angle2} that $\varphi_{\rm L}\left(t\right)$ has a large error at which $\theta_{\rm L}\left(t\right)$ passes by $0$. $\Phi\left(t\right)$ also suffers from the error for the same reason (see Eq.~\eqref{eq:ext_phi}). 
Therefore, the match can be deteriorated if $\theta_{\rm L}\left(t\right)$ passes by $0$ or $\pi$. In fact, in the extracted data in $\theta_{\rm L}\left(0\right)=0$, $\pi/4$ and $\pi$ of SXS:BBH:0037 and $\theta_{\rm L}\left(0\right)=0$ and $\pi$ of SXS:BBH:0164, we find that there is some interval that $\theta_{\rm L}\left(t\right)$ passes by $0$ or $\pi$ during its evolution, and the phase error relative to the QA method increases during this period (see Figs.~\ref{fig:ext_angle2} and~\ref{fig:comp_prec2}). 

Fortunately, the error in $\Phi^{\rm QA}$ would be much smaller than the errors in $\varphi_{\rm L}\left(t\right)$ and $\Phi\left(t\right)$ because these errors are canceled out by taking the combination. For example, for the case that $\theta_{\rm L}\left(t\right)$ is close to $0$, ${\dot \Phi}^{\rm QA}\left(t\right)$ is approximately written as ${\dot \Phi}\left(t\right)+{\dot \varphi}_{\rm L}\left(t\right)$ using Eq.~\eqref{eq:def_psi}. On the other hand, 
$\Phi\left(t\right)+\varphi_{\rm L}\left(t\right)$ is determined only from argument of $h^{\rm ext}_{m=2}\left(t\right)$. Since $h^{\rm ext}_{m=-2}\left(t\right)$ contains the main source of the error in this situation, $\Phi^{\rm QA}\left(t\right)$ is expected to have smaller error than $\varphi_{\rm L}\left(t\right)$ or $\Phi\left(t\right)$. However, as is found in Fig.~\ref{fig:comp_prec2}, some error still remains in $\Phi^{\rm QA}\left(t\right)$, and thus, we still have a room to improve the method for the case that its value passes by $0$ or $\pi$. We note that, for $\theta_{\rm L}\left(0\right)=0$ and $\pi$ of SXS:BBH:0058,  $\theta_{\rm L}\left(t\right)$ also pass by $0$ and $\pi$, respectively. However, the errors in phases are smaller than the cases in SXS:BBH:0037 and SXS:BBH:0164 because the precessing timescale is shorter and the interval staying close to $0$ and $\pi$ are shorter.

	We note that our extraction procedure is only applicable for the case that the $(l,m)=(2,\pm2)$ modes in the co-precessing frame dominates the observed waveforms. While this seems to be a reasonable assumption for the inspiral-stage gravitational waves, to show that this assumption actually holds for the waveforms we employed, we define the relative contribution of the dominant modes for each waveform model by
\begin{align}
	f_{2\pm2}&=\nonumber\\
	&1-\frac{\displaystyle\left<\sum_{m=2,-2}\left|~_{-2}Y^2_{~m}\left[-\theta_{\rm L}\left(t\right),-\psi_{\rm L}\left(t\right)\right]h^{\rm QA}_{2m}\left(t\right)\right|^2\right>}{\displaystyle\left<\sum_{l=2}^{8}\sum_{m=-l}^l\left|~_{-2}Y^l_{~m}\left[-\theta_{\rm L}\left(t\right),-\psi_{\rm L}\left(t\right)\right]h^{\rm QA}_{lm}\left(t\right)\right|^2\right>},\label{eq:domcon}
\end{align}
and summarize $f_{2\pm2}$ for each waveform model in Table~\ref{tb:mismatch}. Here, $\left<\cdot\right>$ denotes the time average over $t=1000\,M$ to the end of the data. We note that $f_{2\pm2}$ depends on the observer direction. Table~\ref{tb:mismatch} shows that the $(l,m)=(2,\pm2)$ modes in the co-precessing frame dominate the observed waveforms for all the waveform models we employed in this paper.

\section{Discussion}
	In this paper, we proposed a new method for extracting the instantaneous orbital axis and for reconstructing the co-precessing waveforms from gravitational waves observed for generic precessing binary black holes. The advantage of our method is as follows: The standard analysis, such as the matched-filter method, requires the template waveform models, in which a particular dynamics of the instantaneous orbital axis is assumed. For example, for black hole-neutron star binaries in close orbits, the orbital precession may not be well described analytically. In such a case, we have an issue for systematically constructing templates. On the other hand, our method does not require a particular model for the dynamics of the instantaneous orbital axis. Thus, it has an advantage to extract the instantaneous orbital axis regardless of its evolution detail; for example, it can be used even in the case that the orbit precesses in a way different from that general relativity predicts as far as the assumptions hold (see the discussions below).
	
	The axis of the precession and the precessing frequency also provide us the information of the total angular momentum of the system for a single spinning binary. The amplitude and the phase of the co-precessing frame waveforms are also reconstructed without modeling their evolutions. Thus, using our method, the co-precessing frame waveforms are direct observables that can be constructed only from detector outputs. The parameter estimation from the precessing waveforms (and the non-precessing waveforms but observed from inclined direction) can be simplified by using the reconstructed co-precessing waveforms since the higher mode templates are not needed or, at least, less needed than using the inertial frame waveforms. In such a case, the number of the template models to cover the parameter space can be reduced by using the approximate mapping between the co-precessing waveforms and non-precessing waveforms~\citep{2012PhRvD..86j4063S}.
	
	There are many other possible applications and extensions for our method. Our method can be extended to extract the higher
modes in co-precessing frame, such as $m=1$ and $m=3$ components. In fact, we find that the co-precessing $(l,m)=(3,\pm3)$ modes agree quite well with the wave components extracted from $|m|=3$.  The amplitude of $m=1$ and $m=3$ modes can be used to solve the degeneracy of parameter estimation between the symmetric mass ratio and the black-hole spin magnitude~\citep{2011PhRvD..84b2002L,2013PhRvD..88f2001A,2013PhRvD..87b4004C,2009PhRvD..80f4027K}. Furthermore, our method can be applied to the data analysis of the waveforms from precessing black hole-neutron star mergers. The previous numerical studies~\citep{2009PhRvD..79d4030S,2010PhRvD..82d4049K,2011PhRvD..84f4018K,2015PhRvD..92h1504P} pointed out that the location of the cutoff in the gravitational wave spectra (the cutoff frequency), caused by tidal disruption of neutron stars, can be used to constrain the neutron star radius. However, the orbital precession (and the inclination of the observer) obscures the location of the cutoff by inducing the modulation in the spectra~\citep{2015PhRvD..92b4014K}. Since this problematic modulation is due to the mixing of the different harmonic components, the method we introduce in this work can be useful to remove such modulation, and may enable us to measure the cutoff frequency in the spectra using the reconstructed co-precessing waveforms~\citep{phdthesis}. These applications and extensions are now in progress.

	 In our method, we made the following assumptions in the analysis: First, we assumed the situation that the complex wave strain is determined with sufficiently high accuracy, and hence, we neglect the effect of the noise and the error of the sky localization for simplicity. Second, we assumed that $(l,m)=(2,\pm2)$ modes of spherical harmonics in the co-precessing frame dominate the strain. Third, the approximate equatorial symmetry is imposed for gravitational waves in the co-precessing frame. The first assumption is made because our purpose is to demonstrate that the direct extraction of the orbital axis and the co-precessing frame waveforms is possible only from the information which we can obtain from the detection in principle. However, of course, data always suffer from the noise in reality. In particular, the sky localization error would be an important source of the error. We should test how well our method works in the presence of the noise and errors, and show what is the required signal-to-noise ratio for achieving the extraction in the required accuracy. The second assumption is made to derive Eqs.~\eqref{eq:phi_orb} and~\eqref{eq:ext_m2}. Although the waveform models we employed in this paper satisfy this assumption (see Table~\ref{tb:mismatch}), those are not still enough to cover the parameter space of the precessing binaries, and we need to extend our exploration to various configurations of precessing binaries; for example, we need to check our method for the case that the ``transitional precession'' occurs~\citep{1994PhRvD..49.6274A}, for which higher-mode contributions to the strain can be significant. The third assumption holds only approximately. As pointed out in Ref.~\citep{2014arXiv1409.4431B}, the equatorial symmetry of the waveforms in the co-precessing frame breaks down in the presence of black hole spin components parallel to the orbital plane. We need to check whether  this assumption is appropriate for the case that the in-plane components are large.

\begin{acknowledgements}
We are grateful to Andrea Taracchini for valuable discussion and checking the paper. We would like to thank the SXS Collaboration for freely providing a variety of high-precision gravitational waveforms of binary-black-hole coalescence. This work was supported by Grant-in-Aid for Scientific Research (Grant Nos. 24244028, 16K05347, 16H02183, 16H06342, 17H01131, 14J02950) of Japanese JSPS. Kyohei Kawaguchi was supported by JSPS Postdoctoral Fellowships for Research Abroad.
\end{acknowledgements}
\bibliographystyle{apsrev4-1}
\bibliography{ref}
\end{document}